\documentclass[aps,prd,superscriptaddress,nofootinbib, preprintnumbers]{revtex4}

\usepackage[english]{babel}
\usepackage[latin1]{inputenc}
\usepackage{hyperref}
\usepackage{graphicx,color}
\usepackage{latexsym}
\usepackage{amsmath}
\usepackage{amsfonts,dsfont}
\usepackage{amssymb,slashed}
\usepackage{bbold}
\usepackage{verbatim,epigraph}
\usepackage{float}

\begin{document}

\title{Orthogonality catastrophe and fractional exclusion statistics}

\author{\textbf{Filiberto Ares }\footnote{ares@unizar.es} }

\affiliation{{Departamento de Fisica Teorica, Universidad de Zaragoza}\\ 
\emph{50009, Zaragoza, Spain}}

\author{\textbf{Kumar S. Gupta}\footnote{kumars.gupta@saha.ac.in} }

\affiliation{{Saha Institute of Nuclear Physics, Theory Division}\\ \emph{1/AF 
Bidhannagar, Kolkata 700 064, India}}

\author{\textbf{Amilcar R. de Queiroz}\footnote{amilcarq@unb.br}}  

\affiliation{{Instituto de
Fisica, Universidade de
Brasilia,} \\ \emph{Caixa Postal 04455, 70919-970, Brasilia, DF, Brazil}}

%\preprint{IMSc/2012/5/12, SINP/TNP/2012/06}

\begin{abstract}
We show that the $N$-particle Sutherland model with inverse-square and harmonic interactions 
exhibits orthogonality catastrophe. For a fixed value of the harmonic coupling, the overlap of 
the $N$-body ground state wave functions with two different values of the inverse-square interaction 
term goes to zero in the thermodynamic limit. When the two values of the inverse-square coupling differ 
by an infinitesimal amount, the wave function overlap shows an exponential suppression. This is qualitatively 
different from the usual power law suppression observed in the Anderson's orthogonality catastrophe.  
We also obtain an analytic expression for the wave function overlaps for an arbitrary set of couplings, whose 
properties are analyzed numerically. The quasiparticles constituting the ground state wave functions of 
the Sutherland model are known to obey fractional exclusion statistics. Our analysis indicates that 
the orthogonality catastrophe may be valid in systems with more general kinds of statistics than just 
the fermionic type.

\end{abstract}
\maketitle

\section{Introduction}

The phenomenon of orthogonality catastrophe (OC) as discussed by Anderson \cite{and} demonstrates that in the thermodynamic limit, the perturbed ground state of certain fermionic quantum systems is orthogonal to the ground state in the absence of the perturbation. The overlap between the two fermionic ground states is usually suppressed by a power law, which goes to zero in the thermodynamic limit. OC has been observed in Kondo systems \cite{k1,k2}, semiconductor quantum dots \cite{k3,k4,k5}, graphene \cite{k6}, Luttinger liquids \cite{ll1,fisher,ll2,ll3,ll4}, and various other physical systems. Recently, a study of statistical OC \cite{sondhi} has led to the possibility of an exponential decay of the wave function overlap \cite{sds}, in 
contrast to the usual power law suppression \cite{and}. A particular way to introduce perturbations in a system is provided by a quench, in which the perturbation could be turned on suddenly or over a small period of time. For a sudden quench, the old ground state is no longer the ground state of the perturbed system, but can be expanded in terms of the complete set of eigenstates of the quenched Hamiltonian. The overlap of the ground states before and after the quench can be used to study the OC. 

The OC has been studied primarily in fermionic systems which obey the Fermi-Dirac statistics. It is therefore interesting to ask if non-fermionic systems, such as those exhibiting fractional exclusion statistics \cite{hs1,hs2,hs3} can also demonstrate OC. In certain fractional quantum Hall systems, where the Laughlin quasiparticles satisfy more general statistics, indirect effects of the OC have been observed mainly through the suppression of the conductance peaks in the thermodynamic limit \cite{wenoc,been}. However, to our knowledge there has been no direct demonstration of OC in terms of the suppression of the wave function overlap for systems with fractional exclusion statistics.

In this paper we want to investigate the existence of the OC in one dimensional quantum systems with fractional exclusion statistics \cite{hs1,hs2,hs3}, 
which is considered as a generalization of the fermionic case. The $N$-body Calogero type systems \cite{calo1,calo2,calo3} with inverse-square and 
harmonic interactions exhibit fractional exclusion statistics \cite{murthy,wadati,poly2,poly3}. The inverse-square interaction is not merely a mathematical curiosity but actually appears in a wide variety of physical situations, including conformal quantum mechanics \cite{cqm,esteve,me-amilcar}, polar molecules \cite{mol1,mol2}, quantum Hall effect \cite{qhe}, Tomonaga-Luttinger liquid \cite{ll}, and black holes \cite{black,danny1,danny2,camb} as well as in graphene with a Coulomb charge  \cite{castro,levi1,levi2,sen1,sen2,sen3,castrormp,wang}. Following the solutions originally obtained by Calogero \cite{calo1,calo2,calo3}, systems with inverse-square interactions have been analyzed with a variety of different techniques \cite{pr,brink1,brink2,kg1,kg2,kg3} and the study of OC with such an interaction is of potential interest for a wide class of physical systems. 

Soon after the appearance of the Calogero model, Sutherland \cite{suth,suth2} proposed a 
variation of that which also exhibits fractional exclusion statistics \cite{poly2,poly3} 
and is more convenient for our purpose. In this paper, we shall use the Sutherland model 
\cite{suth} as a prototype of a quantum system with the inverse-square interaction and the 
harmonic term. The parameters defining the Sutherland model include the inverse-square 
interaction strength $\mu$, the harmonic confining strength $\omega$ and the number $N$ of 
particles that are interacting with each other. We start our analysis with a fixed value 
of $N$ and quench the system parameters from $(\mu, \omega)$ to $(\mu^{\prime}, 
\omega^{\prime})$. The thermodynamic limit will be taken at the end of the calculation. We 
shall show that the overlap of the ground state of the Sutherland model before and after 
the quench decays exponentially in the thermodynamic limit. The wave functions of the 
Sutherland model exhibit fractional exclusion statistics \cite{poly2,poly3}. Therefore the 
results obtained in this paper suggest that the phenomenon of OC might extend to systems 
with statistics more general than just the fermionic type. 

Recent advances in ultra cold atoms and optical lattices have made possible to experimentally realize physical models in lower dimensions. There has been a proposal to experimentally realize the Sutherland model with the help of Bose-Einstein condensates in the cold alkali atoms \cite{cold}. It is therefore plausible that the effects discussed in this paper could be observed in the laboratory in the future.

\section{The Sutherland Model}

The Hamiltonian of the $N$-particle Sutherland model \cite{suth} is given by 
\begin{equation}
\label{ham}
  H_N = \frac{1}{2}\sum_{j=1}^N \Big(-\partial_{x_j}^2 + \omega^2 x_j^2\Big) + 
\sum_{j=2}^N \sum_{k=1}^{j-1}\left(\frac{\mu}{(x_j-x_k)^2} \right), 
\end{equation}
where $\omega$ is a natural frequency common to all the $N$ particles and $\mu\geq 3/4$ is the 
coupling constant for the inverse-square interaction. The ground state wave function for this system has the form
\begin{equation}
\label{hcalo-gs-1} 
\Psi_{\lambda,\omega}\left(\{x\}_N\right)=\mathcal{N}_{(\lambda,\omega)}~z^\lambda~e^{
-\frac { \omega } { 4 }
\sum_{j=1}^N x_j^2},
\end{equation}
where $\lambda\equiv (\sqrt{\mu+1}+1)/2$ is associated with fractional statistics \cite{wadati} and $z\equiv \prod_{j<k}^N (x_j -x_k)$. The normalization constant $\mathcal{N}_{(\lambda,\omega)}$ can be obtained by using the Selberg's integral formula \cite{mehta-book}
\begin{equation}\label{selberg}
 \int_{-\infty}^\infty\cdots \int_{-\infty}^\infty~z^{2\gamma}~\prod_{j=1}^N~e^{-a 
x_j^2}~dx_j=(2\pi)^{N/2}~(2 a)^{-\gamma 
N(N-1)/2-N/2}~\prod_{j=1}^N\frac{\Gamma(1+j\gamma)}{\Gamma(1+\gamma)}.
\end{equation}
Applying (\ref{selberg}) to the ground state (\ref{hcalo-gs-1}), we obtain 
\begin{equation} 
\mathcal{N}_{(\lambda,\omega)}=\left(\frac{\omega}{2\pi}\right)^{N/4}~\omega^{\lambda\frac
{N(N-1)}{4}} ~\prod_ { j=1 } ^N\sqrt{ \frac { 
\Gamma\left(1+\lambda\right)}{\Gamma\left(1+j\lambda\right) } }.
\end{equation}
The dispersion relation of the Sutherland Hamiltonian (\ref{ham}) is that
of $N$ interacting Harmonic oscillators 
\begin{equation}\label{energy}
  E_N=\omega\sum_{i=1}^Nn_i+\omega \lambda\frac{N(N-1)}{2}+\omega \frac{N}{2},
\end{equation}
where $n_i=0,1, 2,\dots$ represents the energy level of the $i$-particle
and satisfying the bosonic occupation rule $n_1\leq n_2\leq \dots \leq n_N$.
Observe that due to the introduction of the inverse-square interaction,
the total energy $E_N$ becomes superextensive, i.e. it is quadratic in
the number of particles $N$.

If we define $\tilde{n}_i=n_i+\lambda(i-1)$, the energy
(\ref{energy}) can be rewritten as
\begin{equation}
  E_N=\omega\sum_{i=1}^N\tilde{n}_i+\omega \frac{N}{2},
\end{equation}
which can be interpreted as the spectrum of $N$ free quasiparticles
obeying the occupation rules
\begin{equation}\label{exclusion}
  \tilde{n}_i\leq \tilde{n}_{i+1}-\lambda.
\end{equation}
They are a generalisation of the Pauli exclusion principle, which corresponds
to the particular case $\lambda=1$. This is consistent with the fact that the
Sutherland model exhibits fractional exclusion statistics \cite{wadati, poly3}.
Note that the superextensive term of $E_N$ in (\ref{energy})
determines the form of the generalised exclusion rules (\ref{exclusion}). 

\section{Scaling of the Overlap between Ground States}

For a fixed value of the particle number $N$, the ground state wave function (\ref{hcalo-gs-1}) of the Sutherland model is characterized by the parameters $\omega$ and $\lambda$. We would like to quench the parameters of this system from $(\lambda,\omega)$ to $(\lambda',\omega')$. Our primary focus is on the strength of the inverse-square interaction term but we also consider the quench in the harmonic interaction as well \cite{Rajabpour}.
To this end, we consider the overlap of the ground states of the Sutherland model with different $(\lambda,\omega)$ and $(\lambda',\omega')$, which is given by
\begin{equation} 
\mathcal{A}_{(\lambda,\omega),(\lambda',\omega')}\equiv 
\Big(\Psi_{(\lambda,\omega)}\left(\{x\}_N\right),\Psi_{(\lambda',\omega')}\left(\{ 
x\}_N\right)\Big)= \int_{-\infty}^\infty\cdots \int_{-\infty}^\infty 
~~\overline{\Psi}_{(\lambda,\omega)}\left(\{x\}_N\right)~\Psi_{(\lambda',\omega')} 
\left(\{x\}_N\right)\prod_{j=1}^N dx_j.
\end{equation}
Using Eq. (\ref{hcalo-gs-1}) in the above formula, we obtain
\begin{equation}
\mathcal{A}_{(\lambda,\omega),(\lambda',\omega')}=\mathcal{N}_{(\lambda,\omega)}~\mathcal{
N } _
{ (\lambda',\omega')}~I\Big(\lambda,\omega;\lambda',\omega'\Big),
\end{equation}
with
\begin{align}
 I\Big(\lambda,\omega;\lambda',\omega'\Big) &=\int_{-\infty}^\infty\cdots 
\int_{-\infty}^\infty 
~z^{\lambda+\lambda'}~\prod_{j=1}^N~e^{-\frac{\omega+\omega'}{4} x_j^2}~dx_j \nonumber \\
    &= (2\pi)^{N/2}~\left(\frac{\omega+\omega'}{2} \right)^{-(\lambda+\lambda') 
N(N-1)/4-N/2}~\prod_{j=1}^N~\frac{\Gamma\left(1+j\frac{\lambda+\lambda'}{2} 
\right)}{\Gamma\left(1+\frac{\lambda+\lambda'}{2}\right)}.
\end{align}
For this evaluation we have used the Selberg's integral formula (\ref{selberg}) to get
\begin{align}\label{overlap-final}
 \mathcal{A}_{(\lambda,\omega),(\lambda',\omega')}&=\left(\frac{4\omega 
\omega'}{(\omega+\omega')^2} \right)^{N/4} \nonumber \\
&\times ~2^{(\lambda+\lambda')N(N-1)/4}~\left(\frac{\omega^\lambda 
\omega'^{\lambda'}}{(\omega+\omega')^{\lambda+\lambda'}}
\right)^{N(N-1)/4} \nonumber \\
&\times 
\left(\frac{\Gamma(1+\lambda)\Gamma(1+\lambda')}{\Gamma(1+(\lambda+\lambda')/2)^2} 
\right)^{\frac{N}{2}} \prod_{j=1}^N~\sqrt{ \frac 
{\Gamma\left(1+j(\lambda+\lambda')/2\right)^2}{ 
\Gamma\left(1+j\lambda\right)\Gamma\left(1+j \lambda'\right)} }.
\end{align}
Notice that the first line gives the overlap due to the harmonic term (see Appendix A), the 
second line mixes both parameters, the natural frequency $\omega$ and the coupling 
$\lambda$, and the third line refers only to the overlap due to the inverse-square interaction strength $\lambda$. Also note that it is not allowed to take the limit $\lambda\to 1$ or equivalently $\mu\to 0$ as we have restricted the inverse-square coupling strength to $\mu \geq 3/4$. Below this value the analysis requires modified boundary conditions related to either self-adjoint extensions \cite{kg1,kg2,kg3} or renormalization of the inverse-square interaction strength \cite{rajeev}. Such modified boundary conditions are not central to the purpose of the present work and that is why we have restricted the analysis to $\mu \geq 3/4$. Moreover, the case $\omega=0$ does not apply in this analysis, where normally the states would be non-normalizable.

For $\lambda=\lambda'$ the harmonic piece remains, while the terms under the product become unity. The mixture piece becomes
 \begin{equation}
  \left(\frac{4 \omega \omega'}{(\omega+\omega')^2} \right)^{\lambda N(N-1)/4},
 \end{equation}
 which can be then combined with the harmonic piece to give
 \begin{equation}\label{overlap-harm}
  \mathcal{A}_{(\lambda, \omega), (\lambda, \omega')} = 
  \left(\frac{4\omega \omega'}{(\omega+\omega')^2}\right)^{N/4+N(N-1)\lambda/4}.
  \end{equation}
In this last case, it is easy to see that the base is positive and smaller than one for 
all $\omega$ and $\omega'$ and the exponent is always positive. Therefore, when 
$N\to\infty$ the whole expression goes to zero in agreement with the OC. The dominant term decays exponentially as 
$\exp(-N^2)$ due to the Calogero coupling. This should be compared with the case of the 
pure harmonic oscillator (see Appendix A) which decays as $\exp(-N)$.

%\begin{figure}[h]
%  \centering
%    \resizebox{14cm}{8cm}{\includegraphics{graph-overlap-1.jpg}}
%    \caption{\textcolor{red}{Some preliminary plots (as a function of $N$), this seems to go to zero. 
%So, there seems to be OC here.}
%}
%  \label{harmonic_loschmidt}
%   \end{figure}

\begin{figure}[H]
  \centering
    \resizebox{15cm}{9cm}{\includegraphics{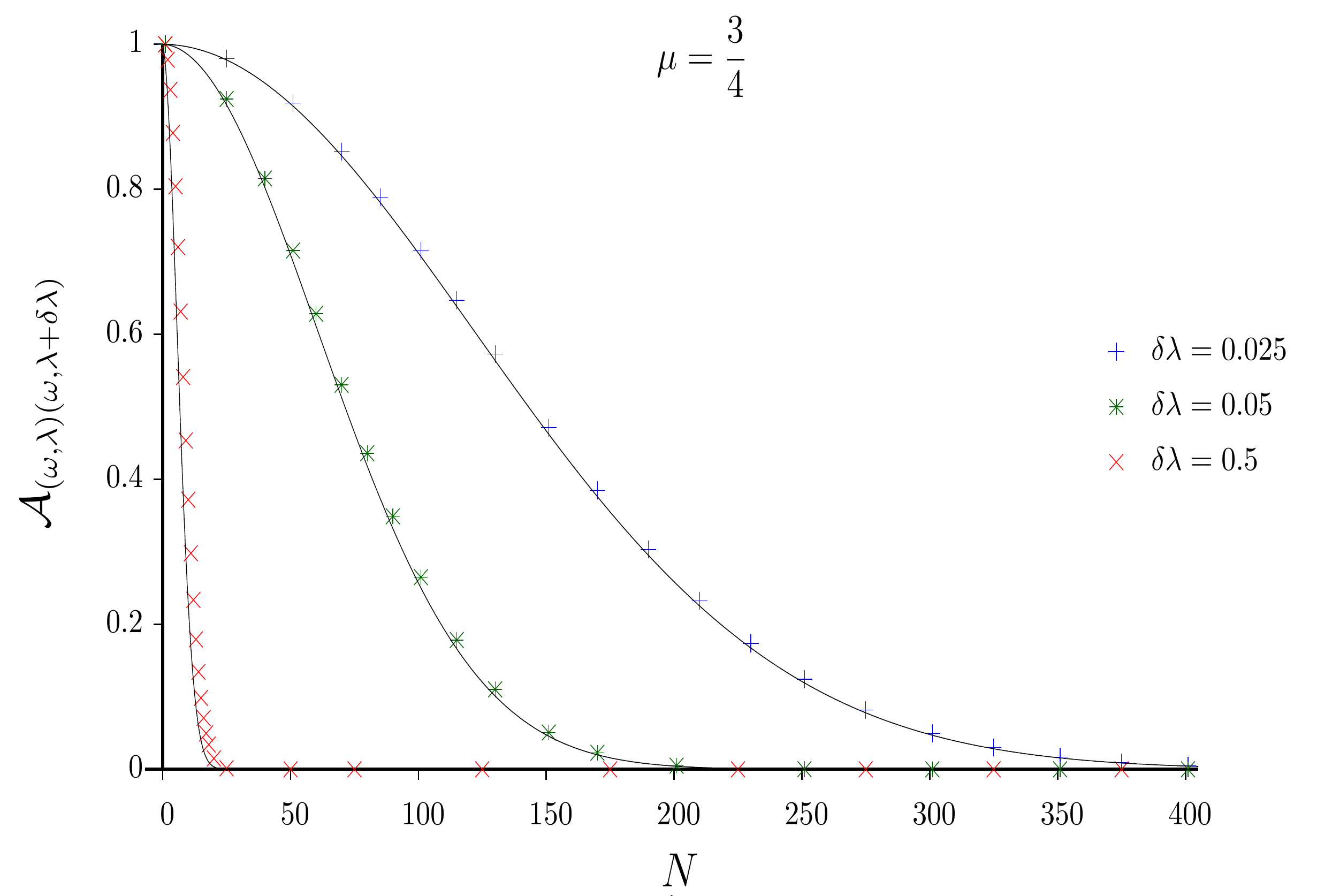}}
    \caption{Overlap  between the ground state of
             a Sutherland model (\ref{ham}) with arbitrary $\omega$ and 
             $\mu=3/4$ and the ground state of another model (\ref{ham}) obtained
             by a perturbation of the inverse-square coupling from 
             $\lambda$ (recall $\lambda= (\sqrt{\mu+1}+1)/2)$) 
             to $\lambda+\delta\lambda$ for different $\delta\lambda$. The harmonic
             term does not change in any case.
             The points were obtained with the exact expression (\ref{overlap-final})
             and the solid lines correspond to the perturbative approximation 
             (\ref{overlap-perturb}). }
  \label{overlap3}
   \end{figure}

We now come to the main result of this analysis, for which we consider the case $\omega=\omega'$. The harmonic piece and the mixed term in (\ref{overlap-final}) become unity. We are now essentially quenching the inverse-square interaction strength from $\lambda$ to $\lambda'$. Following Anderson \cite{and}, let us first consider the case where $\lambda'=\lambda +\delta \lambda$, where  $\delta \lambda \rightarrow 0$ is a small or even infinitesimal perturbation of the inverse-square interaction strength. In this case,  the overlap between the ground states of the initial and the perturbed systems is given by (see Appendix B for details) 
 \begin{equation}\label{overlap-perturb}
 \mathcal{A}_{\lambda,\lambda+\delta\lambda}\sim e^{-\frac{\delta\lambda^2}{\lambda}\frac{N(N+1)}{16}}.
 \end{equation}
Therefore, we find that the overlap exponentially decays to zero with $N^2$, in contrast to the power law suppression as in the Anderson's
original OC \cite{and}. It may be noted that similar exponential suppression has also been recently obtained in a different context \cite{sds}.
In Fig. \ref{overlap3} we compare the perturbative approximation 
(\ref{overlap-perturb}) with the exact expression (\ref{overlap-final}).

We now consider two arbitrarily different values of $\lambda$ and $\lambda'$, 
within the allowed parameter region $\mu \geq 3/4$ and analyze how the overlap 
between the ground states scales with the number of particles $N$ in this case. 
For that, it is useful to apply the Stirling formula for the Gamma function, 
$\Gamma(z)\sim \sqrt{\frac{2\pi}{z}}~z^z~e^{-z}$, when $N\to \infty$ with
$\lambda$, $\lambda'$ keep fixed.  We get
\begin{equation}
 \prod_{j=1}^N\frac{\Gamma\left(1+j\frac{\lambda+\lambda'}{2}\right)}
{\sqrt{\Gamma(1+j\lambda)\Gamma(1+j\lambda')}}\sim 
\left(\frac{(\lambda/2+\lambda'/2)^{\lambda+\lambda'}}
{\lambda^{\lambda}\lambda'^{\lambda'}}\right)^{\frac{N(N+1)}{4}}\left(\frac{\lambda+\lambda'}{2\sqrt{\lambda\lambda'}}\right)^{N/2}.
\end{equation}
The final form of the overlap is obtained as
\begin{equation}\label{overlap-stirling}
  \mathcal{A}_{(\lambda, \omega), (\lambda', \omega)}\sim
  \left(\frac{\Gamma(1+\lambda)\Gamma(1+\lambda')}
  {\Gamma(1+(\lambda+\lambda')/2)^2}\right)^{N/2}
  \left(\frac{((\lambda+\lambda')/2)^{\lambda+\lambda'}}
  {\lambda^\lambda\lambda'^{\lambda'}}\right)^{N(N+1)/4}
  \left(\frac{\lambda+\lambda'}{2\sqrt{\lambda\lambda'}}\right)^{N/2};
 \end{equation}

\begin{figure}[H]
  \centering
    \resizebox{15cm}{9cm}{\includegraphics{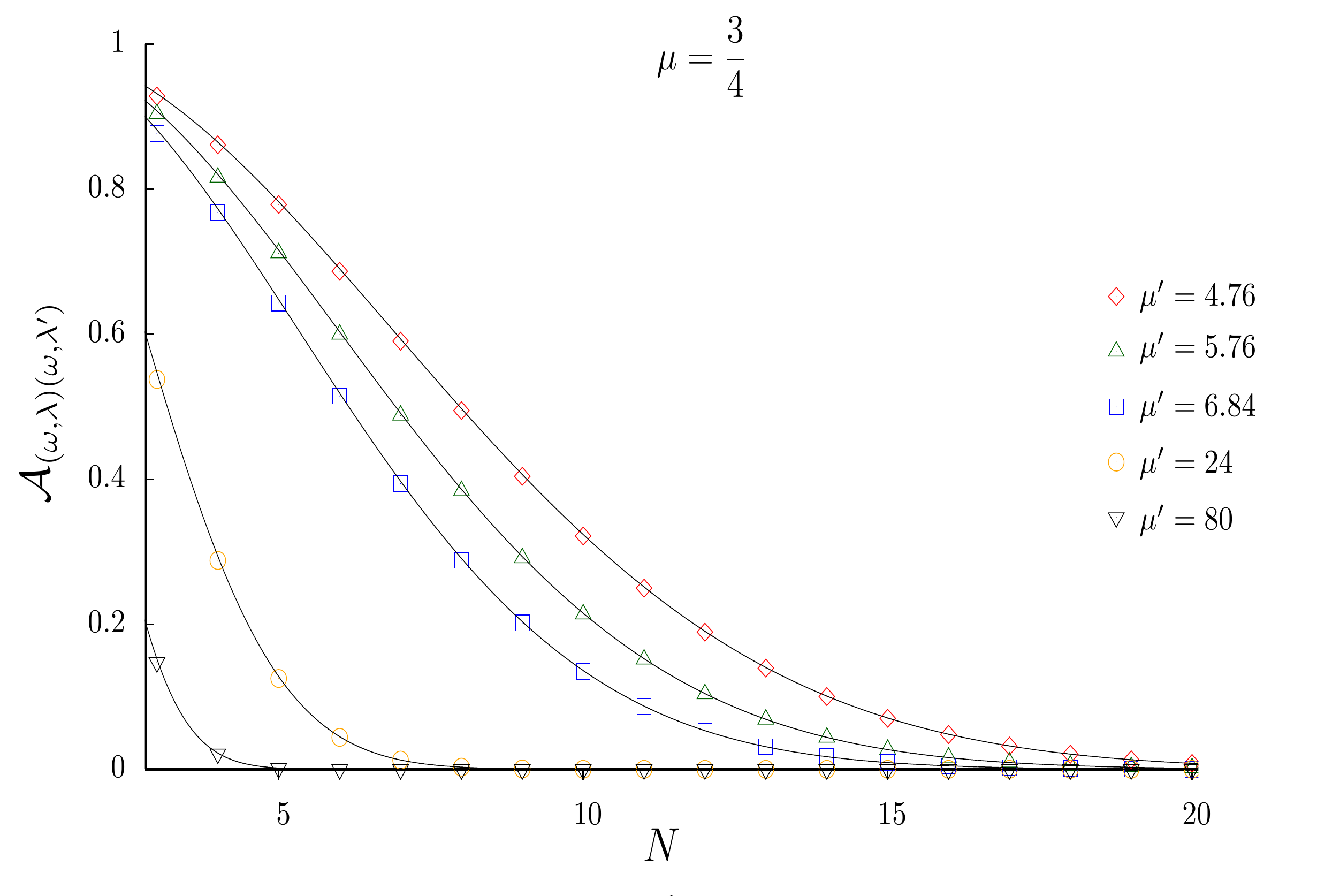}}
    \caption{Overlap between the ground state of
             a Sutherland model (\ref{ham}) with arbitrary $\omega$ and 
             $\mu=3/4$ and the ground state of other 
             Sutherland models (\ref{ham}) with the same Harmonic coupling $\omega$
             and different inverse-square strength $\mu'$. Observe that the
             results do not depend on the value of $\omega$. The points
             were computed employing the exact expression (\ref{overlap-final}) and
             the solid lines correspond to the asymptotic expansion (\ref{overlap-stirling})
             obtained by applying the Stirling formula for the Gamma function when $N\to\infty$.}

  \label{overlap1}
   \end{figure}

In Fig. \ref{overlap1}, we check the validity of this expansion
comparing it with the exact result (\ref{overlap-final})  
for several quenches from a given $\lambda$ to different $\lambda'$.

Observe that, as happens with the energy of the Hamiltonian (\ref{energy}),
the exponent of the previous overlaps is also superextensive,
being quadratic in the number of particles $N$. The inverse-square
interaction seems to be the responsible of this behaviour since
the extensivity is recovered when we turn this interaction off as it is shown in
Appendix A.

%\end{document}

The Sutherland model (SM) can be used to describe a one-dimensional Bose gas in 
an harmonic potential where the particles interact with each other through an inverse-square 
potential. Recent advances in the field of ultra cold atoms and optical lattices have 
opened up the possibility of simulating such a system in the laboratory 
\cite{cold,bhaduri,lazarides, franchini, franchini2}. In particular, it has been argued in Ref. \cite{cold} that the 
dipole-dipole interactions between certain Bose-Einstein condensates (BEC) in an optical 
lattice generates an inverse-square potential whose strength is proportional to the number 
of atoms within the BEC. Thus the coefficient of the inverse-square interaction in the SM 
model can be changed by tuning the number of atoms within the BEC. On the other hand, a 
quench in the harmonic trap of the SM model also leads to interesting 
effects \cite{delcampo} and it can be implemented in the laboratory. 

\section{Summary and outlook}

In our analysis we have considered the $N$-body Sutherland model with the harmonic and 
inverse-square interactions as a prototype for systems with fractional exclusion 
statistics. We have obtained a general analytical expression for the overlap of the ground 
state wave functions when the system parameters are quenched from $(\lambda,\omega)$ to 
$(\lambda',\omega')$. This overlap tends to zero in the thermodynamic limit as the number 
of particles $N \rightarrow \infty$. As a special case, we have considered the quench with 
the harmonic interaction fixed while the inverse-square interaction strength changes 
infinitesimally. Here we have shown that the ground state overlap goes to zero in an 
exponential fashion which is different from the usual power law overlap obtained in the 
usual Anderson's OC. In addition, the exponent is quadratic in the number of particles.
We think that this is a sign of the superextensivity of the Sutherland model, whose dispersion
relation is actually quadratic in the number of particles. This superextensivity
is also related to the fractional exclusion statistics that this model exhibits.
The leading term of the overlap or fidelity of two ground states of a many body system
can be employed to detect and study quantum phase transitions \cite{zanardi, gu}. 
However, we are not aware of any quantum phase transition
in this model. It may be noted that the Calogero-Sutherland system is gapped and there
is no thermal phase transition \cite{suth2}.

Another interesting quantity in this context is the time dependence of the overlap of the wave functions before and after the quench, which  
yields rich information about the equilibrium properties of the quantum system.  The Loschmidt echo \cite{peres,jalabert,physrept} 
\begin{equation}
 L(t) = | \langle \phi_g|e^{iH_it}e^{-iH_ft}|\phi_g\rangle |^2, 
\end{equation}
where $\phi_g$ is the initial state before the quench and $H_i$,$H_f$ denote the Hamiltonians before and after the quench, provides a characterization of such a time dependence and its behavior in the context of Anderson OC has attracted recent attention in the literature \cite{loc1,loc2,loc3,knap,knapreview}. There have also been various proposals to empirically study the Loschmidt echo in setups with ultra cold atoms 
\cite{sds,lexpt1,lexpt2,lexpt3,knap,knapreview} and a Ramsey interferometric type experiment with dilute fermionic impurities has been performed recently \cite{knapexpt}. It is known that the OC is related to the power law decay of the Loschmidt echo \cite{nozieres,knap}. The calculation of the 
Loschmidt echo in the Sutherland model and its analysis in the time and frequency domains is presently under investigation.

\bigskip

\noindent{\bf Acknowledgments:} 
We thank M. Knap for kind comments and for bringing Refs.
\cite{knapreview,knapexpt} to our attention. We also thank A. del
Campo for useful comments, explanations and interest. FA is supported
by FPI Fellowship No. C070/2014, (DGIID-DGA/European Social Fund)
and Grants 2016-E24/2, (DGIID-DGA, Spain), and FPA2015-65745-P
(MINECO, Spain), and thanks the audience and organizers of the
``Martes Cuantico'' seminar at the University of Zaragoza, and
Profs. M. Asorey, J.G. Esteve, F. Falceto, L. Martin-Moreno and D. Zueco
for their interest, comments, and criticism. KSG thanks IIP, Natal, RN,
Brazil for kind hospitality where a part of this work was done. 
ARQ is supported by CNPQ under process number 307124/2016-9.
\appendix 
\section{Harmonic Oscillators}

In this appendix, we recollect a simple example of orthogonality catastrophe. In 
one dimension, consider $N$ non-interacting harmonic oscillators 
\begin{equation}
 H(\omega)=\frac{1}{2}\sum_{j=1}^N~\Big(p_j^2+\omega^2 x_j^2 \Big),
\end{equation}
with $p_j=-i\partial_{x_j}$. The ground state of $H\Psi=E \Psi$ is
\begin{equation} 
\label{gs-harm-1}
\Psi_\omega\left(\{x\}_N\right)=\left(\frac{\omega}{\pi}\right)^{N/4}~e^{-\frac{\omega}{4}
\sum_j^N x_j^2}.
\end{equation}

Let us consider the overlap of two system with different frequencies, $\omega$ and 
$\varpi$,
\begin{equation} 
\mathcal{A}_{\omega\varpi,N}\equiv 
\left(\Psi_\omega\left(\{x\}_N\right),\Psi_\varpi\left(\{ x\} _N\right) \right)=\int 
dx_1\cdots 
dx_N~\overline{\Psi}_\omega\left(\{x\}_N\right)~\Psi_\varpi\left(\{x\}_N\right).
\end{equation}
Replacing (\ref{gs-harm-1}) in the above overlap and integrating the Gaussians,
\begin{equation}
 \mathcal{A}_{\omega\varpi,N} = 
\left(\frac{4\omega\varpi}{(\omega+\varpi)^2}\right)^{N/4}=\left(\frac{4 \eta}{
(1+\eta)^2}\right)^{N/4}, \qquad \eta\equiv \frac{\omega}{\varpi}.
\end{equation}
Observe that the exponent of $\mathcal{A}_{\omega\varpi,N}$ is extensive,
i.e. it depends linearly on the number
of particles $N$. It is clear that the same happens with the total energy
of the system. As we have shown in
the main sections of this work, the inverse-square interaction breaks this
extensivity both at the overlap and at the total energy .

Now, notice that except when $\eta=1$ (or $\omega=\varpi$), the ratio $4\eta/(1+\eta)^2$ 
is less than one. Therefore, in the limit $N\to\infty$,
\begin{equation}
 \lim_{N\to \infty}\mathcal{A}_{\omega\varpi,N} = 0, \qquad \omega\neq \varpi.
\end{equation}
This is the orthogonality catastrophe. It is a remarkable fact that the study of the 
scaling of this exponential suppression is currently amenable to experiments in cold 
atoms.

\section{Perturbative analysis for the OC of the Sutherland Model}

For the case $\omega'=\omega$, the overlap between two ground states 
(\ref{hcalo-gs-1}) with $\lambda$ and $\lambda'$ as a function of 
the number of particles $N$ writes from (\ref{overlap-final}) as
\begin{equation}
\label{overlap-1}
 \mathcal{A}_{\lambda,\lambda'}(N) = R_1(\lambda,\lambda')^\frac{N}{2} 
\sqrt{\prod_{j=1}^N \frac{1}{R_j(\lambda,\lambda')}}=\exp\left(\frac{N}{2}\log 
R_1(\lambda,\lambda') - \frac{1}{2}\sum_{j=1}^N \log R_j(\lambda,\lambda') \right),
\end{equation}
where
\begin{equation}
 R_j(x,y)\equiv \frac{\Gamma(1+jx) \Gamma(1+jy)}{\Gamma\left(1+j(x+y)/2 \right)^2}.
\end{equation}

We are interested in the asymptotics of the overlap function 
$\mathcal{A}_{\lambda,\lambda'}(N)$ as $N\to\infty$. We will also consider a small
perturbation around $\lambda$, that is, $\lambda'=\lambda +\delta \lambda$. The
result below will be valid for the scaling limit
\begin{equation}
 N\to\infty,\,\, \delta \lambda \to 0, \qquad N\delta\lambda^2  <1.
\end{equation}

The series expansion around $x$ of $R_j(x,x+\delta x)$ 
to smallest order is 
\begin{equation}
\label{Taylor-Series-1}
 R_j(x,x+\delta x)=1+j^2 \psi'(1+j x) \frac{\delta x^2}{4} + \mathcal{O}(\delta x^3),
\end{equation}
where  
\begin{equation}
 \psi'(z)=\sum_{k=0}^\infty \frac{1}{(k+z)^2}, \qquad z\neq 0,-1,-2,\dots,
\end{equation}
is the derivative of the Polygamma function
\begin{equation}
 \psi(z)=\frac{\Gamma'(z)}{\Gamma(z)}.
\end{equation}
Notice that the argument in $\psi'$ for our case 
is always positive. The asymptotics of $\psi'$ as $z\to\infty$ 
is of the form
\begin{equation}
 \psi'(z)\sim \frac{1}{z}+\frac{1}{2z^2}+\sum_{k=1}^\infty \frac{B_{2k}}{z^{2k}},
\end{equation}
where $B_{n}$ are the Bernoulli numbers.

Now we consider the case $j\to\infty$, and since $x$ is held fixed, then we can take
\begin{equation}
 \psi'(1+j x) \sim \frac{1}{j\left(x+\frac{1}{j}\right)}\sim\frac{1}{jx}.
\end{equation}
Replacing this result in (\ref{Taylor-Series-1}), we obtain
\begin{equation}\label{expand_r}
 R_j(x,x+\delta x)\sim 1+ j\frac{\delta x^2}{4 x}.
\end{equation}

Using the scaling limit $N\delta x^2<1$, for large $N$ and small $\delta x$, we then 
Taylor expand the previous formula to first order as
\begin{equation}\label{log_expand_r}
 \log R_j(x,x+\delta x) \sim \log \left(1+ j\frac{\delta x^2}{4 x} \right)
 = j\frac{\delta x^2}{4 x}.
\end{equation}
We can also consider that formulae (\ref{expand_r}) 
and thus (\ref{log_expand_r}) are  valid for $j=1$. 

We now replace this formula in the sum over $j$ in (\ref{overlap-1}). Although the above 
formula is valid 
for large $j$, we can start the sum with $j=1$ which brings nothing more than a small 
error. Indeed, the smaller $j$ terms are quite irrelevant w.r.t. the large ones (in a 
linear approximation). Thus,
\begin{equation}
 \sum_{j=1}^N \log R_j(x,x+\delta x) \sim  \frac{\delta 
x^2}{4 x}\sum_{j=1}^N j =  \frac{\delta 
x^2}{4 x} \frac{N(N+1)}{2}.
\end{equation}

Summing up all the above in the overlap expression (\ref{overlap-1}), we obtain
 \begin{equation}
 \mathcal{A}_{\lambda,\lambda+\delta \lambda}(N) \sim ~e^{-\frac{\delta 
\lambda^2}{\lambda}\frac{N(N+1)}{16}}.
\end{equation}

\end{document}